\documentclass[]{interact}

\usepackage[caption=false]{subfig}

\usepackage[numbers,sort&compress,merge]{natbib}
\bibpunct[, ]{[}{]}{,}{n}{,}{,}
\renewcommand\bibfont{\fontsize{10}{12}\selectfont}

\theoremstyle{plain}

\theoremstyle{definition}

\theoremstyle{remark}

\usepackage{graphicx}
\usepackage{dcolumn}
\usepackage{bm}

\usepackage{tabularx}

\usepackage{xcolor}
\definecolor{rojo}{RGB}{181,72,45}     
\definecolor{magenta}{RGB}{112,40,190} 
\definecolor{azul}{RGB}{44,92,167}     
\definecolor{verde}{RGB}{117,120,26}   
\definecolor{blanco}{RGB}{255,255,255} 
\definecolor{cafe}{RGB}{43,17,0}

\definecolor {MyColor1}    {RGB} {44,92,167}
\definecolor {MyColor2}    {RGB} {44,92,167}
\usepackage{amsmath}
\usepackage{colortbl}

\usepackage{hyperref}
\hypersetup{
    citecolor=azul,
    colorlinks=true,
    linkcolor=azul,
    filecolor=magenta,      
    urlcolor=azul
}

\usepackage[flushleft]{threeparttable}
\usepackage{footnote}

\usepackage{float}
\usepackage{tikz}
\usepackage{tikz-3dplot}
\usetikzlibrary{babel,arrows,intersections,shapes.arrows,fadings}
\usepackage{pgfplots}
\pgfplotsset{compat=newest}             
\usepgfplotslibrary{fillbetween}
\usepackage{filecontents}

\begin{document}

\articletype{ARTICLE TEMPLATE}

\title{Molecular Nature of Thermodynamics and its Application to Nanosize Particle Systems}

\author{
\name{Eduardo Hern\'andez-Huerta\textsuperscript{a}, 
Ruben Santamaria\textsuperscript{a}\thanks{CONTACT Ruben Santamaria Author. Email: rso@fisica.unam.mx} 
and Tom\'as Rocha Rinza\textsuperscript{b}}
\affil{\textsuperscript{a} Departamento de F\'isica Te\'orica, Instituto de F\'isica, UNAM, M\'exico; \\ 
\textsuperscript{b} Departamento de F\'isicoqu\'imica, Instituto de Qu\'imica, UNAM, M\'exico.}
}

\maketitle

\begin{abstract}
\noindent The origins of thermodynamics from the microscopic properties of matter have not been satisfactorily 
accounted for. This work presents a formulation that connects Lagrangian mechanics to thermodynamics. By 
using such a formulation and following similar steps to those performed in the laboratory, the heat capacities, 
energetic and mechanical response coefficients, absorbed and emitted heats, entropy changes, and 
thermodynamic energies of a prototype water system, employed as an illustrative example, are found to be 
close to the experimental results on water bulk. The present formulation is realistic. After connecting Lagrangian mechanics 
to Langevin stochastic dynamics, the first law of thermodynamics is derived. The formulation not only exhibits 
the transformation of time-reversible equations onto time-irreversible equations, and the molecular 
origins of heat as well, but also reveals new pathways to investigate the thermodynamics of systems of 
nanometric sizes.
\end{abstract}


\section {Introduction} \label{ththti}

\noindent Thermodynamics is the field of science that deals with the different forms of energy, and their transformations. Thermodynamics is well consolidated, and has 
been extended to include the different states of matter, with broad applications in the industry and technology. Yet, it is regarded as a scientific branch with a 
primordial experimental character. On the other side, Lagrangian mechanics is the field of science that deals with the mechanical and dynamical aspects of classical 
objects. The modern mechanical formulations lack of providing a rational basis to thermodynamics and, in consequence, the separation between Lagrangian mechanics and 
thermodynamics remains in our time.

\medskip

\noindent The laws of thermodynamics represent the pillars of thermodynamics. For example, the \emph{Zero Law of Thermodynamics} establishes the heat flux from hot 
bodies to cold bodies to reach the thermal equilibrium, and the \emph{First Law of thermodynamics} states that every physical system has internal energy that can be 
changed by the action of external factors, like mechanical work and heat. This law is different to the mechanical law of energy conservation due to the presence of the 
heat term in the expression of the first law. The thermodynamic laws have empirical character \cite{Mandl1988}. Yet, we know that they have a microscopic origin, and 
there should be an explanation from a more fundamental theory \cite{Salinger1982}. The first attempts to link the microscopic theory to the macroscopic theory are 
attributed to the creators of kinetic theory \cite{Mikhailov2005, Maxwell1860, Naim2010}. In fact, they established the foundations of \emph{Statistical Mechanics}, 
where statistical approaches are combined with (classical and quantum) mechanics to determine the properties of the macroscopic bodies with ensemble averages.

\medskip

\noindent The modern simulation methods propose a Lagrangian with a bigger number of degrees of freedom than those required to describe the particle system alone, because 
the particle system is in contact with an environment \cite{Hover1985}. The additional degrees of freedom may be visualized as fictitious particles capable of interacting 
with the real particles \cite{Andersen1980, Rondoni2002}. It is through the presence of the fictitious particles that temperature and pressure effects are considered in 
the particle system \cite{Nose1984a, Nose1984b}. The simplifications introduced by the extended Lagrangian approaches brings some issues, and it has not been possible to 
deduce the first law of thermodynamics from them. A different perspective departs from the Langevin approach \cite{Sekimoto2010, Mohammad2017, Seifert2012}. A connection 
between the deterministic mechanical dynamics and classical thermodynamics has not been fully achieved. An alternative form to establish the connection between 
microscopic and macroscopic properties employs ensemble theory \cite{Esposito2015}. It is focused in a kind of modified version of thermodynamics, which goes from the 
macroscopic to the microscopic behaviors of the particles \cite{Hill1994}. In short, we have formulations that in going from the microscopic domain to the macroscopic 
one, and vice versa, lack of a consistent connection between the microscopic and macroscopic regimes \cite{Haddad2017}.

\medskip

\noindent The goal of this work is to provide an approach different to the ones currently known in the literature \cite{Terrell2001, Dick2020} to close gaps between 
Lagrangian mechanics and thermodynamics. To do this, a thermodynamic prototype constituted by particles is proposed (section \ref{thpatms}), and described using 
fundamental mechanics. Thus, the Lagrangian and motion equations are given in section \ref{tlagpro}. Mechanical properties such as the temperature, volume, total energy, 
and more, are presented in sections \ref{mechtemp} and \ref{totene}, with results of the thermodynamic observables shown in section \ref{thoaps}. Section \ref{conclusions} 
presents the conclusions of this work. The appendixes provide details of the model interaction potentials employed in this work.

\section {Microscopic thermodynamic prototype} \label{thpatms}

\begin{figure}[t]
\centering
\includegraphics [width=0.9\textwidth,keepaspectratio] {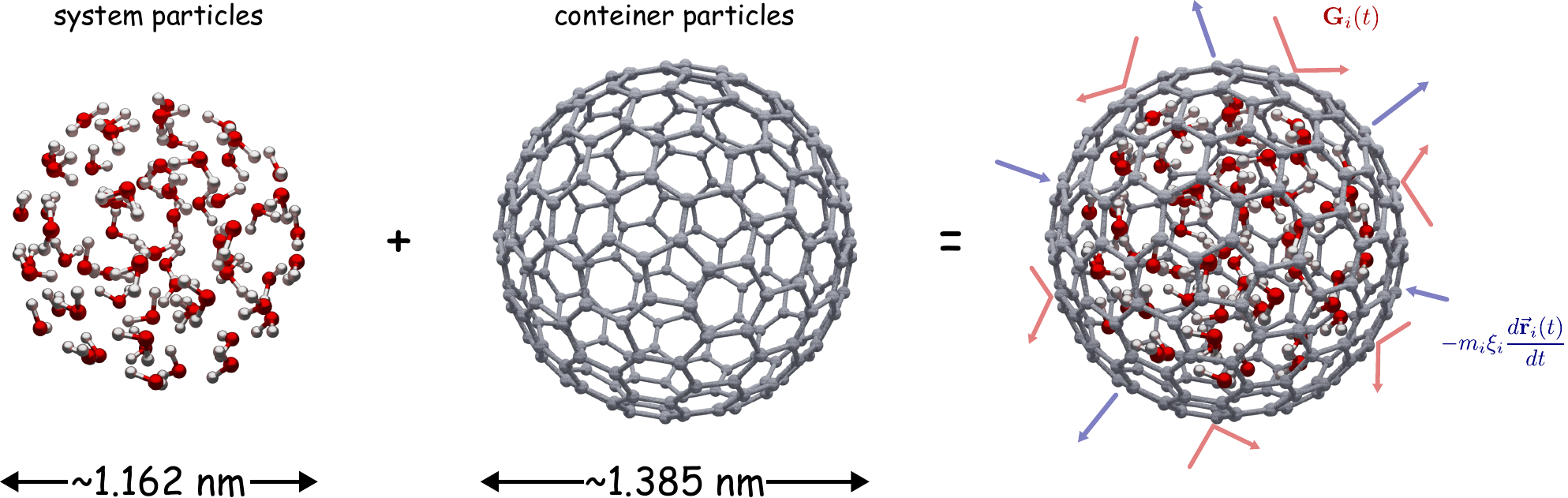}
\caption {The microscopic prototype is characteristic of a macroscopic thermodynamic system. It is integrated by three sub-systems: $(i)$ the (relevant) set of particles 
to be confined and investigated, $(ii)$ the particles that build the container, and $(iii)$ the particles of the heat reservoir, which are not depicted. Yet, the 
molecular formulation derived from the microscopic model is realistic, as the equations have universal character, and engulf the thermodynamic description of macroscopic 
systems.} \label{fig.01}
\end{figure}

\noindent The elements of the microscopic model are characteristic of a general (macroscopic) thermodynamic system. Such elements are: $(i)$ the system of particles 
(atoms or molecules) in confinement, $(ii)$ the container, whose atomic structure is considered in this approach, and $(iii)$ the thermal reservoir, whose atomic 
structure is also considered and has interaction with the container. The system of confined particles is the relevant system of particles since we require to establish 
the thermodynamics of this system. There is no restriction on the container's geometrical shape and size. The thermal reservoir is supposed to have macroscopic dimensions. 
The nomenclature to distinguish the types of particles is: ``$s$'' (from soup) for the particles under confinement, ``$x$'' for the particles building the container, 
and ``$q$'' for the particles of the environment. The molecular system to investigate is water under confinement due to the vast information we have on this system in 
the literature.  The number of water molecules to deal with is 64 under ambient conditions of temperature $(\sim 25^\circ {\rm C})$, pressure $(\sim 1\, {\rm atm})$, 
and density (1.0\, g/mL). They are confined in a container of fullerene type integrated by 240 carbons, with a radius defined by the water density under ambient 
conditions. The container is immersed in a thermal reservoir (Fig. \ref{fig.01}). Although the microscopic model has specific structure, geometrical shape, and size, the 
equations to determine in the following sections have universal nature, as they are not restricted to a given number of atoms, type of atom, or specif form of atomic 
arrangement. Yet, the proposed microscopic model allows to carry molecular simulations with the present computers. The following sections present the theoretical 
formulation and, to some extent, follow the work of Santamaria et al. \cite{Ruben2016}. The final sections provide the numerical results on the thermodynamics 
observables in a way of proof of principle.

\section {The Lagrangian proposal} \label{tlagpro}

\noindent The thermodynamic model involves the confined particles (also referred to as the \emph {system particles}), the particles of the container, and those of the 
heat reservoir. The Lagrangian is a functional of the individual degrees of freedom of such particles \cite{Goldstein1981}:

\vspace{-0.35cm}
\begin{align}
\displaystyle \mathcal L (s_k, x_i, q_{ij}, \dot s_k, \dot x_i, \dot{q}_{ij}, t)=& \sum_i \frac {m_i} 2 \dot x_i^2 + \sum_{ij} \frac {m_{ij}} 2 \dot {q}_{ij}^2 + \sum_i \frac {m_k} 2 \dot s_k^2 \\[-0.1cm]
 &\hspace{-3.0cm}  -E_{qx} (\{ q_{ij}, x_i \})-E_{xs} (\{ x_i, s_k \}) -E_{ext} (\{ x_i, s_k \}) \nonumber
\end{align}
\vspace{-0.5cm}

\noindent The positions $s_k$, $x_i$ and $q_{ij}$ correspond to the confined particles, container particles, and heat-bath particles, respectively. Their masses are 
$m_k$, $m_i$, and $m_{ij}$, where the indexes refer to individual particles. The mass $m_{ij}$ carries two indexes as the $j$th particle of the thermal bath is in 
interaction with the $i$th particle of the container ($m_{ij}$ should not be confused with the reduced mass). The potential energy $E_{qx}$ provides the interactions 
among the heat-bath particles (of type $q-q$), and of such particles with the container particles (of type $q-x$). The potential energy $E_{xs}$ gives the interactions 
among the container particles (of type $x-x$), of such particles with the confined particles (of type $x-s$), and the interactions among the confined particles (of 
type $s-s$). The potential energy $E_{ext}$ represents an external field acting on the confined particles and the container particles. 

\medskip

\noindent The individual particles of the thermal reservoir play a second role in this investigation because they do not interact directly with the confined system of 
particles. Therefore, it is convenient to introduce approximations of the energy term $E_{qx}$. The first step in this quest is to consider that each particle $x_i$ has 
interaction with its own set of $q$-particles. This is a mathematically important step because it helps to uncouple the interactions among the container particles in the 
equations of the heat-bath particles. Yet, the separation of the heat-bath particles in sets for the $x$-particles is alleviated later. The harmonic approximation is 
used to describe the interaction of the thermal-reservoir particles with those of the container \cite{Caldeira1983, Haake1984}:

\vspace{-0.45cm}
\begin{equation} \label{tlagpro.2}
\displaystyle E_{qx}= \sum_{ij} \left[ \frac {m_{ij} \omega_{ij}^2} 2\, q_{ij}^2 -c_{ij} q_{ij} x_i +\ \frac {c_{ij}^2} {2m_{ij} \omega_{ij}^2}\, x_i^2 \right]
\end{equation}
\vspace{-0.2cm}

\noindent The harmonic potential $E_{qx}$ considers the individual $q$ and $x$ vibrational degrees of freedom, including the coupling terms $q_{ij} x_i$ with strength 
$c_{ij}$, and the vibrational frequencies $\omega_{ij}$ \cite{Ullersma1966}. The energy term $E_{xs}$ is approximated in the framework of molecular-mechanics with a 
forcefield, but quantum-mechanical calculations can be similarly performed:

\vspace{-0.8cm}
\begin{equation} \label{tlagpro.3}
\displaystyle E_{xs}= U_x^{str} + U_{xs}^{vdw} +\ U_s^{el} + U_s^{vdw} + U_s^{str}
\end{equation}
\vspace{-0.5cm}

\noindent $U_x^{str}$ is a potential of harmonic type for the ($x-x$ bonded) interactions of the container particles, $U_{xs}^{vdw}$ represents the van der Waals ($vdw$) 
($x-s$ non-bonded) interactions of the container particles and the confined particles, $U_s^{el}$ and $U_s^{vdw}$ stand for the Coulombic and $vdw$ ($s-s$ non-bonded) 
interactions of the confined particles, and $U_s^{str}$ is a potential of harmonic type for the ($s-s$ bonded) interactions of the individual molecules that constitute 
the confined particles. The details of the different interaction models are found in Appendix A.

\subsection {The equations of motion} \label{teqomot}

\noindent The Euler-Lagrange method gives the procedure to obtain the motion equations from the Lagrangian. The equations describe the time evolution of the three types 
of particles:

\vspace{-0.7cm}
\begin{equation} \label{teqomot.2}
\begin{tabular} {l}
$\displaystyle m_{ij} \ddot {q}_{ij}= c_{ij} x_i -m_{ij} \omega_{ij}^2 q_{ij}$\\[0.25cm]

$\displaystyle m_i \ddot x_i= \sum_j\bigg( c_{ij} q_{ij} -\frac {c_{ij}^2} {m_{ij} \omega_{ij}^2}\, x_i \bigg) -\frac {\partial E_{xs}} {\partial x_i} -\frac {\partial E_{ext}} {\partial x_i} $\\[0.45cm]

$\displaystyle m_k \ddot s_k= -\frac {\partial E_{xs}} {\partial s_k} -\frac {\partial E_{ext}} {\partial s_k}$
\end{tabular}
\end{equation}
\vspace{-0.25cm}

\noindent It is possible to solve the first expression analytically due to the uncoupling of the $x_i$ particle from the other particles of type $x_j$ ($j \ne i$) in the 
equations of the $q_{ij}$ particles. The solution is inserted into the second expression \cite{Zill2016, Zwanzing2001}. It can be shown that the container particles are 
influenced by two main terms: a memory kernel $K_i$ and a noise source $G_i$ \cite{Ruben2016}. In spite of that, the solutions for the container particles are still 
mathematically complicated to obtain. In order to reach the solutions, the kernel term is simplified by assuming a heat bath of weakly-coupled interacting oscillators, 
statistically described by the Debye theory. A short-time memory kernel leads to a dissipative force acting on the $i$th atom of the container, and the stochastic term 
$G_i$ is determined by a Chandrasekhar's bivariate distribution function of the position and velocity of the $i$th atom \cite{Chandrasekhar1943}. The distribution 
function alleviates the initial supposition on the existence of different sets of heat-bath particles for the $x_i$ atoms of the container, because a single distribution 
function is applied to all the container atoms. The final expressions show the coupling of the Newtonian and Langevian equations, including the heat-bath effects in 
statistical manner:

\vspace{-0.5cm}
\begin{equation} \label{teqomot.3}
\begin{aligned}
{\rm Langevin:} & \quad m_i \ddot x_i= -\frac {\partial E_{xs}} {\partial x_i} - \frac {\partial E_{ext}} {\partial x_i} -m_i \xi_i\, \frac {dx_i (t)} {dt} + G_i (t) \\[0.1cm]
{\rm Newton:}   & \quad m_k \ddot {s_k}= -\frac {\partial E_{xs}} {\partial s_k} -\frac {\partial E_{ext}} {\partial s_k}
\end{aligned}
\end{equation}
\vspace{-0.2cm}

\noindent The Langevin viscosity $\xi_i$ and temperature $T$ are the parameters that characterize the reservoir. The terms $-m_i \xi_i dx_i/ dt$ and $G_i$ represent 
accelerating and decelerating terms connected to the fluctuation-dissipation effects of the heat bath, and acting on the container particles. The last expression of Eqs. 
(\ref{teqomot.3}) describes the dynamics of the confined particles in interaction with the container particles. It keeps the deterministic nature of the Newtonian 
expression because we are interested on determining the thermodynamic state of the confined particles. It is important to remark that other approaches of the literature 
include fictitious forces on the confined particles via extended degrees of freedom in the Lagrangian, and are combined with periodic boundary conditions. The sampling of 
the phase space using those approaches is expected to differ from the sampling obtained with the present approach due to the existence of fictitious forces and the use 
of periodic boundary conditions in the other approaches. The predictions of such approaches are improved when scale factors are applied to the ensemble averages, with 
the purpose of compensating for the action of the fictitious forces on the real particles. In this work, there are no scale factors as the confined particles are 
subjected to real forces, this is, every confined particle only interacts with neighbor particles and the container.

\medskip

\noindent The original Lagrangian equations of motion are time-symmetric. Nevertheless, the equations of motion of the container atoms evolved onto stochastic expressions, 
in turn affecting the dynamics of the confined particles. In this regard, the trajectories of the confined particles in the backwards time direction are not expected to 
reproduce the trajectories of the confined particles obtained in the forward time direction. This is due to the action of the stochastic forces on the container 
particles, which interact with the confined particles. The stochastic terms summarize the effects of a macroscopic heat bath, composed by many particles. Thus, those 
terms bring some issues in the formulation, like the lost of information on the system, which is the cause for no reproducing the particle trajectories discussed 
previously. However, this is a general feature of any formulation that introduces simplified terms, in particular, the thermodynamic theory. Finally, the set of 
Eqs. (\ref{teqomot.3}) may be also used to describe a body with atomic structure moving stochastically. This is, a \emph{Brownian body}, whether it is microscopic or 
macroscopic in size, with any geometrical shape. In the limit of reducing the Brownian body to a single particle, it is possible to reproduce the stochastic dynamics of 
a single particle \cite{Sekimoto1998}.
\vspace{-0.3cm}

\section {Mechanical temperature, confinement volume, and particle density} \label{mechtemp}

\noindent A \emph {mechanical temperature}, $T_{mec}$, is defined in terms of the average kinetic energy of the confined particles \cite{Tildesley2017}:

\vspace{-0.29cm}
\begin{equation} \label{mechtemp.1}
\begin{tabular} {c}
$\displaystyle T_{mec}= \frac 2 {(3N-6) k_B} \sum_k \frac {m_k} 2\, \dot {\bf s}\,_k^2$
\end{tabular}
\end{equation}
\vspace{-0.3cm}

\noindent The number of degrees of freedom of the confined particles is $3N-6$, and $k_B$ is the \emph{Boltzmann constant}. Let us also consider the statistical 
temperature $T_{st}$ imposed in the laboratory through the heat bath. The thermal equilibrium is achieved when $T_{mec}= T_{st}$. In this regard, the 
\emph {Zero Law of Thermodynamics} is satisfied. The simulations satisfy the \emph {ergodic hypothesis} when: $T_{mec}= T_{st}$, and the particle dynamics allows to 
cover the entire phase-space that the confined particles are entitled to. In our example, the container has spherical shape, and the confined volume is:

\vspace{-0.5cm}
\begin{equation} \label{confvol.1}
\displaystyle 
V_{conf}= \frac 4 3 \pi R_{conf}^3 \qquad \Rightarrow \qquad
R_{conf}= R_{part} + \frac 1 2 (R_{cage} -R_{part})
\end{equation}
\vspace{-0.35cm}

\noindent The confining radius $R_{conf}$ is given in terms of the radii $R_{part}$ and $R_{cage}$, where $R_{part}$ is the radius of the most external particles in 
confinement, and $R_{cage}$ is the average radius of the container particles. The amount $[R_{cage} -R_{part}]/ 2$ is the correction factor that considers the occupied 
volume by the electron cloud of the most external confined particles. This last contribution is important in systems of nanometric size, becoming less important in 
macroscopic systems. The density of the confined system is $\rho= N/ V_{conf}$, in particle/ \AA$^3$. To compare the particle density with the bulk density, we use 
conversion factors to express $\rho$ in g/mL:

\vspace{-0.5cm}
\begin{equation} \label{sysdens.1}
\begin{tabular} {c}
$\displaystyle \rho= \frac {N M_w} {N_A V_{conf}} \times 10^{24}$ \hspace{0.3cm} [g/mL]
\end{tabular}
\end{equation}
\vspace{-0.35cm}

\noindent $M_w$ is the mass in grams of a mole of particles, $N_A$ is the Avogadro's number, and $V_{conf}$ is the confinement volume in \AA$^3$.
\vspace{-0.29cm}

\section {Total energy} \label{totene}

\noindent The present approach introduces stochastic terms in the equations of motion of the container atoms, and the total energy should be computed accordingly. To do 
this, the equations of motion are transformed onto energy equations \cite{Alonso1967}:

\vspace{-0.4cm}
\begin{equation} \label{totene.1}
\begin{tabular} {rl}
$\displaystyle m_i \ddot {\bf x}_i \cdot d{\bf x}_i $&$\displaystyle = -\frac {\partial E_{xs}} {\partial \bf x_i} \cdot d{\bf x}_i -\frac {\partial E_{ext}} {\partial {\bf x}_i} \cdot d{\bf x}_i - m_i \xi_i\, \frac {d{\bf x}_i (t)} {dt} \cdot d{\bf x}_i + {\bf G}_i (t) \cdot d{\bf x}_i $\\\\[-0.25cm]


$\displaystyle m_k \ddot {\bf s}_k \cdot d{\bf s}_k$&$\displaystyle= -\frac {\partial E_{xs}} {\partial {\bf s}_k} \cdot d{\bf s}_k -\frac {\partial E_{ext}} {\partial {\bf s}_k} \cdot d{\bf s}_k$
\end{tabular}
\end{equation}
\vspace{-0.3cm}

\noindent It is convenient to separate the \emph{conservative forces} (with nomenclature $f$) from the \emph{non-conservative forces} (with nomenclature $g$):

\vspace{-0.2cm}
\begin{equation} \label{totene.2}
\begin{tabular} {c}
$\displaystyle {\bf f}_{x_i}^{int}= -\frac {\partial E_{xs}} {\partial {\bf x}_i} \quad ; \quad {\bf f}_{s_k}^{int}= -\frac {\partial E_{xs}} {\partial {\bf s}_k} \quad ; \quad  {\bf f}_{x_i}^{ext}= -\frac {\partial E_{ext}} {\partial {\bf x}_i} \quad ; \quad {\bf f}_{s_k}^{ext}= -\frac {\partial E_{ext}} {\partial {\bf s}_k}$ \\\\[-0.2cm]


$\displaystyle {\bf g}_{x_i}= -m_i \xi_i\, \frac {d{\bf x}_i (t)} {dt} + {\bf G}_i (t)$
\end{tabular}
\end{equation}
\vspace{-0.1cm}

\noindent When the elements of Eqs. (\ref{totene.1}) are added, and integrated, we obtain:

\vspace{-0.3cm}
\begin{equation} \label{totene.5}
\begin{tabular} {r}
$\displaystyle \sum_i \int_{\dot {\bf x}_i^{(0)}}^{\dot {\bf x}_i} m_i \dot {\bf x}_i \cdot d \dot {\bf x}_i + \sum_k \int_{\dot {\bf s}_k^{(0)}}^{\dot {\bf s}_k} m_k \dot {\bf s}_k \cdot d \dot {\bf s}_k=$\\\\[-0.2cm]

$\displaystyle \sum_i \int_{{\bf x}_i^{(0)}}^{{\bf x}_i} {\bf f}_{x_i}^{int} \cdot d{\bf x}_i + \sum_k \int_{{\bf s}_k^{(0)}}^{{\bf s}_k} {\bf f}_{s_k}^{int} \cdot d{\bf s}_k$\\\\[-0.2cm]

\hspace{0.0cm} $\displaystyle + \sum_i \int_{{\bf x}_i^{(0)}}^{{\bf x}_i} {\bf f}_{x_i}^{ext} \cdot d{\bf x}_i + \sum_k \int_{{\bf s}_k^{(0)}}^{{\bf s}_k} {\bf f}_{s_k}^{ext} \cdot d{\bf s}_k$\\\\[-0.2cm]

$\displaystyle + \sum_i \int_{{\bf x}_i^{(0)}}^{{\bf x}_i} {\bf g}_{x_i} \cdot d{\bf x}_i$
\end{tabular}
\end{equation}
\vspace{-0.2cm}

\noindent The integrals involving the conservative forces are due to an ordered work, $W$, while the non-conservative forces characterize a \emph{disordered work}, which 
is an energy in transition and is recognized as \emph{heat}, $Q$:

\vspace{-0.6cm}
\begin{equation} \label{totene.6}
\begin{tabular} {c}
$\displaystyle W_x^{int}= \sum_i \int_{{\bf x}_i^{(0)}}^{{\bf x}_i} {\bf f}_{x_i}^{int} \cdot d{\bf x}_i \qquad ; \qquad W_s^{int}= \sum_k \int_{{\bf s}_k^{(0)}}^{{\bf s}_k} {\bf f}_{s_k}^{int} \cdot d{\bf s}_k $\\\\[-0.2cm]

$\displaystyle W_x^{ext}= \sum_i \int_{{\bf x}_i^{(0)}}^{{\bf x}_i} {\bf f}_{x_i}^{ext} \cdot d{\bf x}_i \qquad; \qquad\ W_s^{ext}= \sum_k \int_{{\bf s}_k^{(0)}}^{{\bf s}_k} {\bf f}_{s_k}^{ext} \cdot d{\bf s}_k$\\\\[-0.25cm]

$\displaystyle Q= \sum_i \int_{{\bf x}_i^{(0)}}^{{\bf x}_i} {\bf g}_i \cdot d{\bf x}_i = \sum_i \int_{{\bf x}_i^{(0)}}^{{\bf x}_i} \left( -m_i \xi_i\, \frac {d{\bf x}_i(t)} {dt} + {\bf G}_i(t) \right)  \cdot d{\bf x}_i $
\end{tabular}
\end{equation}
\vspace{-0.25cm}

\noindent Solving the integrals on the left side of Eq. (\ref{totene.5}), and inserting expressions (\ref{totene.6}) in Eq. (\ref{totene.5}), gives:

\vspace{-0.35cm}
\begin{equation} \label{totene.9}
\begin{tabular} {c}
$\displaystyle \sum_i \left\{ \frac {m_i} 2 \dot {\bf x}_i^2 -\frac {m_i} 2 \left[ \dot {\bf x}_i^{(0)} \right]^2 \right\} +\ \sum_k \left\{ \frac {m_k} 2 \dot {\bf s}_k^2 -\frac {m_k} 2 \left[ \dot {\bf s}_k^{(0)} \right]^2 \right\}=$\\\\[-0.25cm]

$\displaystyle W_x^{int} + W_s^{int} + W_x^{ext} + W_s^{ext} + Q$
\end{tabular}
\end{equation}
\vspace{-0.3cm}

\noindent The internal works of Eq. (\ref{totene.9}) are defined by the difference of a potential energy function evaluated at the initial and final states:

\vspace{-0.45cm}
\begin{equation} \label{totene.10}
W^{int}_x= E_{p,x}^{(0)} -E_{p,x} \quad ; \quad W^{int}_s= E_{p,s}^{(0)} -E_{p,s}
\end{equation}
\vspace{-0.35cm}

\noindent By using these energy differences in Eq. (\ref{totene.9}), and replacing the quantity $E_{p,x} + E_{p,s}$ by the function $E_{xs}$, we obtain:

\vspace{-0.2cm}
\begin{equation} \label{totene.12}
\begin{tabular} {c}
$\displaystyle \sum_i \frac {m_i} 2\, \dot {\bf x}_i^2 + \sum_k \frac {m_k} 2\, \dot {\bf s}_k^2 + E_{xs} (\{ {\bf x}_i, {\bf s}_k\}) - $\\\\[-0.25cm]

$\displaystyle \bigg\{ \sum_i \frac {m_i} 2 \left[ \dot {\bf x}_i^{(0)} \right]^2 + + \sum_k \frac {m_k} 2 \left[ \dot {\bf s}_k^{(0)} \right]^2 +\ E_{xs} (\{ {\bf x}_i^{(0)}, {\bf s}_k^{(0)} \}) \bigg\} $\\\\[-0.25cm]

$= W_x^{ext} + W_s^{ext} + Q$
\end{tabular}
\end{equation}
\vspace{-0.4cm}

\noindent The terms that define the \emph{total internal energy}, $E_{int}$, of the whole particle system are:

\vspace{-0.4cm}
\begin{equation} \label{totene.13}
\begin{tabular} {c}
$\displaystyle E_{int}= \sum_i \frac {m_i} 2\, \dot {\bf x}_i^2 + \sum_k \frac {m_k} 2\, \dot {\bf s}_k^2 + E_{xs}$
\end{tabular}
\end{equation}
\vspace{-0.38cm}

\noindent The final expression with the new nomenclature is:

\vspace{-0.55cm}
\begin{equation} \label{totene.14}
\displaystyle \Delta E_{int} = E_{int} -E_{int}^{(0)} = W_x^{ext} + W_s^{ext} + Q
\end{equation}
\vspace{-0.6cm}

\noindent Eq. (\ref{totene.14}) is the \emph{First Law of Thermodynamics}. Sekimoto carried out an analysis using a single Langevian particle and obtained a first-law 
like energy expression \cite{Sekimoto1997, Sekimoto1998}. That work is important as it illustrates a link between thermodynamics and stochastic dynamics. In our work we 
observe the same type of connection for the particles forming the container. Nevertheless, our analysis additionally includes terms of a confined system whose description 
is, in fact, fundamental since it includes no stochastic terms. In this form, it is possible to connect the thermodynamic theory to stochastic dynamics, and this last 
one to the deterministic Lagrangian dynamics of the confined system of particles.

\section {Energy of confinement} \label{teneoco}

\noindent It is important to obtain the \emph{internal energy}, $U$, of the confined particles. Thus, the energy of the container particles is subtracted from $E_{int}$ 
presented in Eq. (\ref{totene.13}):

\vspace{-0.4cm}
\begin{equation} \label{teneoco.2}
\begin{tabular} {l}
$\displaystyle  U= E_k +  E_p \quad ; \quad E_k= \sum_k \frac {m_k} 2\, \dot {\bf s}_k^2$\\\\[-0.25cm]

$E_p= E_{xs} -U_x^{str} = U_{xs}^{vdw} + U_s^{el} + U_s^{vdw} + U_s^{str}$
\end{tabular}
\end{equation}
\vspace{-0.25cm}

\noindent The above internal energy is the \emph{confining energy}. It is not conserved because the contact of the confined particles with the container brings the 
\emph{thermal fluctuations} into consideration. The thermal fluctuations are significant at the microscopic level and less significant at the macroscopic one 
\cite{Bustamante2005}.
\vspace{-0.2cm}

\section {Parametrization of the confining energy} \label{parotce}

\noindent It is convenient to have an analytic expression of the confining energy $U$ in terms of $N$, $V$, $T$. To do this, we require to carry a number of simulations 
using different confining volumes, and maintaining the temperature fixed: $(E_p^{(1)},V_1,T)$, $(E_p^{(2)},V_2,T)$, $\cdots$, $(E_p^{(n)},V_{n},T)$. Several isotherms are 
generated by changing the temperature. The Vinet curve is adequate in this context \cite{Vinet1986}:

\vspace{-0.5cm}
\begin{equation}\label{parotce.1}
\displaystyle E_p (V,T)= a(T) + b(T)\, e^{\gamma(T)\, V^{1/3}} + c(T)\, V^{1/3}\, e^{\gamma(T)\, V^{1/3}}
\end{equation}
\vspace{-0.6cm}

\noindent The parameters $a$, $b$, $c$, and $\gamma$ are obtained after fitting the Vinet curve to the required isotherm \cite{Santamaria2012}. Under such conditions, an 
analytical internal energy of the confined particles is achieved:

\vspace{-0.6cm}
\begin{equation} \label{parotce.3}
\begin{tabular} {c}
$\displaystyle U (N,V,T)= \frac 1 2 (3N-6) k_BT + E_p (V,T)$
\end{tabular}
\end{equation}

\section {Mechanical pressure} \label{mechpre}

\noindent The pressure registered on the confined particles is a mechanical pressure, $P_{mec}$, with static and dynamic contributions:

\vspace{-0.6cm}
\begin{equation} \label{mechpre.1}
\begin{tabular} {rl}
$\displaystyle P_{mec}= P_s + P_d= -\frac {\partial E_p (V,T_{eq})} {\partial V} + \frac {Nk_B T_{eq}} V$
\end{tabular}
\end{equation}
\vspace{-0.35cm}

\noindent Eq. (\ref{mechpre.1}) can use the analytic form of $E_p (V,T)$ given in Eq. (\ref{parotce.1}):

\vspace{-0.4cm}
\begin{equation} \label{mechpre.2}
P_{mec} (N,V,T)\displaystyle = \frac {Nk_B T_{eq}} V -\Big( \frac {\gamma b} {3V^{2/3}} + \frac {\gamma c} {3V^{1/3}} +\ \frac c {3V^{2/3}} \Big) \, e^{\gamma V^{1/3}}
\end{equation}
\vspace{-0.35cm}

\noindent The system pressure may consider a reference pressure  $P= P_{mec} -P_0$.

\section {Thermodynamics of a particle system} \label{thoaps}

\noindent One of the advantages of the present formulation resides on the fact that we can follow similar steps to those performed in the laboratory for the thermodynamic 
characterization of the confined particle system. The challenge is to work in a kind of worst-case scenario, where the system to deal with is composed by a finite number 
of particles, far from the thermodynamic limit, and the particle interactions are described by effective potentials (Appendix A). The work on such a system and the 
realization of such steps may be also understood as a proof of principle of the present formulation.

\section {Simulation of water in the liquid state} \label{sowitls}

\noindent The temperature $T=$ 300\, K and viscosity coefficient $\xi=$ 5\, ps$^{-1}$ of the heat-bath fluid are defined before starting the simulations. The molecular 
model with 64 confined water molecules and a spherical container consisting of 240 carbon atoms is employed. The simulations demand average simulation times of 200\, ps, 
with an integration step of 1\, fs. The mechanical temperature $T_{mec}$ is corroborated with the statistical temperature $T_{st}$ of the heat bath, and the averages of 
the physical variables are calculated in 50\, ps intervals. The radial distribution function of the oxygen atoms shows structural O$\cdots$O peaks at 2.8 \AA\ and 
3.5 \AA. The experimental measurements show peaks at 2.8--2.9 \AA, characterizing the first-solvation shell, and a plateau between 3.5 \AA\ and 4.5 \AA, characterizing 
the second-solvation shell \cite{Ricci2000, Nilsson2015}. The model using a finite number of water molecules shows to be consistent with the structure of the macroscopic 
water bulk.

\medskip

\noindent As an observation, we find that the number of particles forming the container plays a second role in determining the internal energy of the confined particles. 
Furthermore, the ratio $U_{xs}^{vdw}/ U$ decreases when the number of confined particles increases, as expected.

\section {Constant volume process} \label{convolpro}

\begin{figure} [!t]
\centering
\includegraphics [width=0.45\textwidth,keepaspectratio] {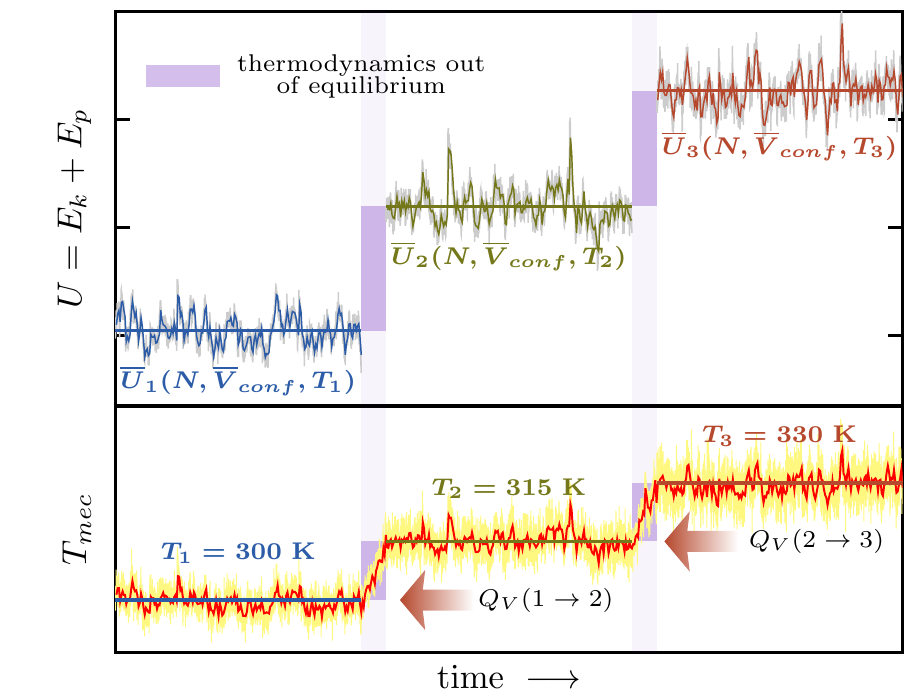}
\caption {$(i)$ The instantaneous internal energy and $(ii)$ instantaneous mechanical temperature are shown. The heat-reservoir temperature is 300, 315, and 330\, K for 
the system of 64 confined water molecules under constant volume. The values of the average energies $\overline U_i$ are given in {\color{MyColor1} Table \ref{table.01}}.} \label{fig.02}
\end{figure}

\noindent The heat capacity at constant volume is calculated with reservoir temperatures of 300, 315 and 330\, K, maintaining the number of particles and volume fixed. 
The fluctuations of the temperature and internal energy are presented in {\color{MyColor2} Fig. \ref{fig.02}}. The internal energy $U$ is given by the addition of the 
kinetic energy $E_k$ and potential energy $E_p$, Eqs. \ref{teneoco.2}. The horizontal lines in {\color{MyColor2} Fig. \ref{fig.02}} represent the time averages of the 
physical variables. The average energies are provided in {\color{MyColor1} Table \ref{table.01}}. The values are similar to those using different types of methodologies 
in the literature \cite{Ramirez2009}.

\medskip

\begin{table}[b]
\tbl {Time averages of the internal energy and its energy components using different equilibrium temperatures.$^\ast$}
{\begin{tabular}{ccccccc} \toprule
$\overline T_{mec}$ [K] && $\overline E_p$ [au] && $\overline E_k$ [au] && $\overline U$ [au]\\
(Eq. \ref{mechtemp.1}) && (Eq. \ref{teneoco.2}) && (Eq. \ref{teneoco.2}) && $\overline E_k + \overline E_p$\\
\cline{1-1} \cline{3-3} \cline{5-5} \cline{7-7}
$299.4$ && $-0.9558$ && $0.2702$ && $-0.6856$\\
$315.2$ && $-0.9345$ && $0.2845$ && $-0.6500$\\
$330.3$	&& $-0.9206$ && $0.2981$ && $-0.6225$\\ \bottomrule
\end{tabular}}
\tabnote{
$^\ast$The particle density is 1 g/mL.
}\label{table.01}
\end{table}

\begin{table}
\tbl {Time average properties of 64 water molecules confined in a 240 carbon-atom container.$^\ast$}
{\begin{tabular}{cccccccccccccc} \toprule
$\overline V_{conf}$ & $\rho$ & \multicolumn {4} {c} {$T=$ 300 K} && \multicolumn {3} {c} {$T=$ 315 K} && \multicolumn {3} {c} {$T=$ 330 K}\\
\cline{4-6} \cline{8-10} \cline{12-14}
$[{\rm nm}^3]$ & [g/ml] &&  $\overline E_p$ & $\overline U$ & $P$ && $\overline E_p$ & $\overline U$ & $P$ &&
$\overline E_p$ & $\overline U$ & $P$\\
\hline
1.9070 & 1.00 && -0.9558 & -0.6856 &      1.0 && -0.9345 & -0.6500 &    302.4 && -0.9206 & -0.6225 &    426.9 \\
1.8383 & 1.04 && -0.9498 & -0.6774 &  2,182.4 && -0.9324 & -0.6471 &  2,587.5 && -0.9100 & -0.6113 &  2,738.1 \\
1.7230 & 1.11 && -0.9285 & -0.6561 &  8,525.2 && -0.9069 & -0.6217 &  9,226.2 && -0.8868 & -0.5877 &  9,448.2 \\
1.6178 & 1.18 && -0.8829 & -0.6110 & 19,841.2 && -0.8574 & -0.5735 & 21,057.6 && -0.8429 & -0.5455 & 21,399.5 \\
1.5193 & 1.26 && -0.7970 & -0.5260 & 40,169.0 && -0.7797 & -0.4953 & 42,290.2 && -0.7554 & -0.4575 & 42,836.1 \\ \bottomrule
\end{tabular}}
\tabnote{ $^\ast$The units of the potential energy $\overline E_p$, Eq. (\ref{parotce.1}), and the internal energy $\overline U$, Eq. (\ref{parotce.3}), are au, and of the 
pressure $P$, Eq. (\ref{mechpre.2}), are atm. The average kinetic energies evaluated at different temperatures are $E_k$ (300 K)= 0.2716 au, $E_k$ (315 K)= 0.2847 au, 
and $E_k$ (330 K)= 0.2982 au. The average densities, $\rho$, and volumes of confinement, $\overline V_{conf}$, are also reported.
}\label{table.02}
\end{table}

\noindent The \emph{caloric curve} is obtained from the previous data. According to the temperature range given in {\color{MyColor1} Table \ref{table.01}}, the 
energy $U$ is linearly related to the temperature, $U(T)= 2.03962 \times 10^{-3}\, T -1.29514$ au. The differentiation of this expression with respect to $T$ gives the 
heat capacity $C_V$. The computed value is $2.03962 \times 10^{-3}$ au/K. The equivalent specific value is 4.6485 J/ (g K). The experimental value is 4.0893 J/ (g K). 
The computed values are in reasonable agreement with the experiment. Other theoretical values of $C_V$ reported in the literature show larger deviations from the 
experimental values \cite{Paesani2006, Shiga2005}.

\medskip

\noindent The heat transferred to the confined system of particles to produce the temperature changes given in {\color{MyColor1} Table \ref{table.01}} is computed from 
$Q_V= \Delta U$, the \emph{mechanical equivalent of heat}. In the change of states 300 K $\rightarrow$ 315 K, and 315 K $\rightarrow$ 330 K, the transferred heat is 
1.4604 kJ/mol and 1.1281 kJ/mol, respectively. Such data is close to the experimental values 1.1294 kJ/mol and 1.1292 kJ/mol under the same temperature changes if we 
consider that the investigated case is considered a worst-case scenario.

\section {Constant pressure process} \label{conprepro}

\begin{figure} [!b]
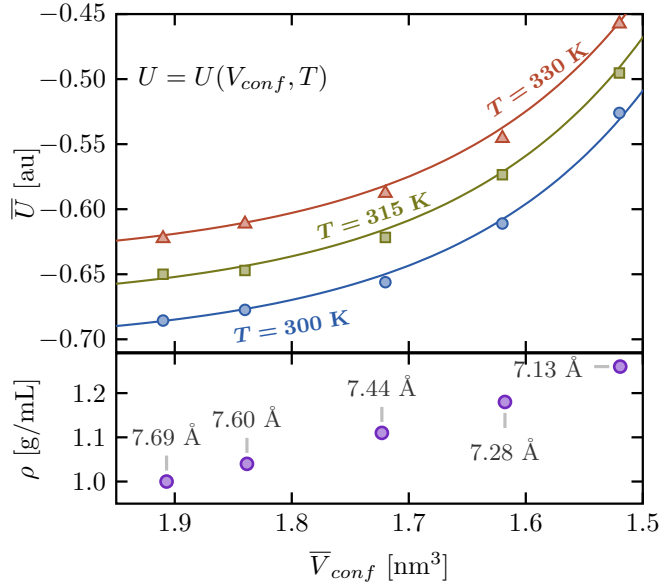

\centering
\input{plot/data/energy-vol}
\input{plot/data/vol-rho}

\pgfmathdeclarefunction{eVinet}{0}{\pgfmathparse{\a + \b*exp(\d*x^(1/3)) + \c*x^(1/3)*exp(\d*x^(1/3)) + \ekin}}
\begin{tikzpicture}[scale=1.0, font=\small]
\begin{axis}[name=main plot,scale only axis, 
	     width=7.0cm,  height=4.5cm,
	     ylabel={$\overline{U}\ [\rm au]$},  y label style={at={(axis description cs:-0.125,.5)},rotate=0},
	     ymin=-0.71, ymax=-0.45, ytick distance=0.05,
	     xmin=1.5, 	xmax=1.95, 	xtick distance=0.1, x dir=reverse,
	     xticklabel=\empty,
	     line width=0.4pt, 
	     axis line style={black, line width=0.30mm}, 
	     every tick label/.append style={black},
	     every tick/.style={line width=0.30mm},
	     every axis label/.append style ={black},
	     yticklabel=\pgfkeys{/pgf/number format/.cd,fixed,precision=2,zerofill}\pgfmathprintnumber{\tick},
	     ]
	     
\node[fill=none,rotate=0] at (1.85,-0.50) {\textcolor{black}{$U = U(V_{conf},T)$}};
\def\a{-0.9787}
\def\b{6.55e+10}
\def\c{6.74e+09}
\def\d{-23.2678}
\def\ekin{0.2713}
\addplot[
    domain=1.15:2.35,
    samples=200,
    color=azul,
    name path=f,
    line width=0.80pt
    ]
    {eVinet};
\addplot+[only marks, mark=*,  
          azul, 
          mark options={fill=azul!50, solid}, 
          mark size=2.0pt, line width=0.75pt
         ] table[x index=0,y index=1] {energy-vol.dat};
\node[fill=none,rotate=9] at (1.8,-0.69) {\footnotesize\textcolor{azul}{$\boldsymbol{T = 300\ {\rm K}}$}};
\def\a{-0.9605}
\def\b{6.55e+10}
\def\c{6.74e+09}
\def\d{-23.2274}
\def\ekin{0.2847}
\addplot[
    domain=1.15:2.35,
    samples=200,
    color=verde,
    name path=f,
    line width=0.80pt
    ]
    {eVinet};
\addplot+[only marks, mark=square*,
          verde, 
          mark options={fill=verde!50, solid}, 
          mark size=1.8pt, line width=0.75pt
         ] table[x index=0,y index=2] {energy-vol.dat};
\node[fill=none,rotate=17] at (1.73,-0.605) {\footnotesize\textcolor{verde}{$\boldsymbol{T = 315\ {\rm K}}$}};
\def\a{-0.9412}
\def\b{6.50e+10}
\def\c{6.74e+09}
\def\d{-23.2123}
\def\ekin{0.2983}
\addplot[
    domain=1.15:2.35,
    samples=200,
    color=rojo,
    name path=f,
    line width=0.80pt
    ]
    {eVinet};
\addplot+[only marks, mark=triangle*,  
          rojo, 
          mark options={fill=rojo!50, solid}, 
          mark size=2.8pt, line width=0.75pt
          ] table[x index=0,y index=3] {energy-vol.dat};
\node[fill=none,rotate=34] at (1.59,-0.50) {\footnotesize\textcolor{rojo}{$\boldsymbol{T = 330\ {\rm K}}$}};
\end{axis}

\begin{axis}[at={(main plot.below south west)}, anchor=north west, yshift=0.25cm,
             visualization depends on = \thisrow{angle} \as \angle,
	     width=7.0cm, scale only axis,  height=2.0cm,
	     ylabel={$\rho\ [\rm g/mL]$}, xlabel={$\overline{V}_{conf}~[\rm nm^3]$}, 
	     ymin=0.95, ymax=1.29, ytick distance=0.1, 
	     xmin=1.5, xmax=1.95, 	xtick distance=0.1, x dir=reverse,
	     line width=0.4pt, 
	     axis line style={black, line width=0.30mm}, 
	     every tick label/.append style={black},
	     every axis label/.append style ={black},
	     every tick/.style={line width=0.30mm},
	     y label style={at={(axis description cs:-0.125,.5)},rotate=0}, 
	     yticklabel=\pgfkeys{/pgf/number format/.cd,fixed,precision=1,zerofill}\pgfmathprintnumber{\tick},
    	     xticklabel=\pgfkeys{/pgf/number format/.cd,fixed,precision=1,zerofill}\pgfmathprintnumber{\tick},
	     ]     
\addplot+[  only marks, mark=*,  magenta, mark options={fill=magenta!50, solid}, mark size=0.80mm, line width=0.35mm,
            nodes near coords=,
            point meta = explicit symbolic,
            every node near coord/.style = {
            anchor = center, pin = {[pin edge={gray!50,line width=1.2pt},pin distance=0.2cm] \angle:\pgfplotspointmeta}
            },
            every node/.style={font=\fontsize{9.0}{0}\selectfont,text=black!80},
         ] table[col sep = space,meta=label] {vol-rho.dat};
\end{axis}
\end{tikzpicture}  
\caption {The average internal energy, Eq. (\ref{teneoco.2}), and the confinement volume, Eq. (\ref{confvol.1}), are shown along the isotherms 300 K, 315 K, 330 K. 
The numerical data are represented by points. The curves were generated analytically after: $(i)$ using Eq. (\ref{parotce.1}) in Eq. (\ref{teneoco.2}) for the potential 
energy $E_p$, and $(ii)$ fitting Eq. (\ref{parotce.1}) to the $E_p$ data. The density and confinement radius, Eq. (\ref{confvol.1}), of the confined water molecules is 
also given for the different data points.} \label{fig.03}
\end{figure}

\noindent The calculation of the heat capacity at constant pressure, $C_P$, is difficult as the enthalpy is required, and the enthalpy demands knowledge of the pressure 
$P$ at every equilibrium state. In order to evaluate $C_P$, a series of simulations with fixed temperature $T$, and changing the volume $V_{conf}$, are performed to 
simulate an \emph {isothermal compression process}. Several isotherms of the type $E_p= E_p (V, T=$ constant) are determined, and shown in {\color{MyColor1} Table 
\ref{table.02}}. The use of Eq. (\ref{parotce.1}) allows to parametrize $E_p$ to reproduce the data of {\color{MyColor1} Table \ref{table.02}}. The values of the 
parameters $a$, $b$, $c$, and $\gamma$ of Eq. (\ref{parotce.1}) are given in {\color {MyColor1} Table \ref{table.03}}. After the insertion of Eq. (\ref{parotce.1}) 
in Eq. (\ref{parotce.3}), we obtain the analytic expressions of the isotherms $U_1= U (V,T_1= 300\, {\rm K}$), $U_2= U (V,T_2= 315\, {\rm K}$) and 
$U_3= U (V,T_3= 330\, {\rm K}$) ({\color {MyColor2} Fig. \ref{fig.03}}).

\begin{table}[t]
\tbl {Parameters of the potential energy, Eq. (\ref{parotce.1}), in terms of the temperature.} 
{\begin {tabular} {ccccccccc} \toprule
$T$   && $a$    &&  $b/10^{10}$ && $c/10^9$ && $\gamma$\\
$[{\rm K}]$ && $[{\rm au}]$ &&  $[{\rm au}]$      && $[{\rm au}/{\rm nm}]$  && $[1/{\rm nm}]$\\
\cline{1-1} \cline{3-3} \cline{5-5} \cline{7-7} \cline{9-9}
300 && -0.9787 && 6.55 && 6.74 && -23.2678\\
315 && -0.9605 && 6.55 && 6.74 && -23.2274\\
330 && -0.9412 && 6.55 && 6.74 && -23.2123\\ \bottomrule
\end{tabular}}
\label{table.03}
\end{table}

\medskip

\begin{table}[b]
\tbl {The confinement volume, $V_{conf}$, internal energy, $U$, and enthalpy, $H$, under different conditions of temperature and pressure.$^\ast$}
{\begin {tabular} {ccccccccccccc} \toprule
&& \multicolumn {3} {c} {$V_{conf}\, [{\rm nm}^3]$ (Eq. \ref{confvol.1})}
&& \multicolumn {3} {c} {$U$ [au] (Eq. \ref{parotce.3})}
&& \multicolumn {3} {c} {$H$ [au] $= U + PV_{conf}$}\\
\cline{3-5} \cline{7-9} \cline{11-13}
$T_{eq}$ [K] && 1 atm & 50 atm & 100 atm && 1 atm & 50 atm & 100 atm && 1 atm & 50 atm & 100 atm\\
\cline{1-1} \cline{3-5} \cline{7-9} \cline{11-13}
300 && 1.907 & 1.905 & 1.903 && $-0.6854$ & $-0.6852$ & $-0.6850$ && $-0.6853$ & $-0.6830$ & $-0.6805$\\
315 && 1.918 & 1.917 & 1.915 && $-0.6543$ & $-0.6541$ & $-0.6538$ && $-0.6542$ & $-0.6518$ & $-0.6494$\\
330 && 1.923 & 1.921 & 1.919 && $-0.6216$ & $-0.6214$ & $-0.6212$ && $-0.6216$ & $-0.6192$ & $-0.6167$\\ \bottomrule
\end{tabular}}
\tabnote{
$^\ast$The pressure is normalized to $P_0$ (given in the body text). Once the pressure $P$ and temperature $T$ are fixed, the confinement volume $V_{conf}$ is 
determined, and the internal energy is evaluated using Eq. (\ref{parotce.3}). Up to four digits after the decimal point are presented with the purpose of observing 
differences in the energies.
}
\label{table.04}
\end{table}

\noindent The mechanical pressure is analytically computed from Eq. (\ref {mechpre.2}). Firstly, it is necessary to provide a reference pressure. The liquid state of 
water at 300\, K and 1\, atm is taken as a reference state and, thus the difference $P= P_{mec} -P_0$ is calibrated to give 1\, atm for the system with density 
1.0\, g/mL at 300\, K. The pressure values are reported in {\color{MyColor1} Table \ref{table.02}}, with $P_0=$ 6,070.3\, atm.

\begin{figure} [t]
\centering
\pgfmathdeclarefunction{pVinet}{0}{\pgfmathparse{(-exp(\d*x^(1/3))*((\d*\b)/(3*x^(2/3))+(\d*\c)/(3*x^(1/3))+\c/(3*x^(2/3))))*(1e27*4.3597e-18)/101325 + (0.138065/1.01325)*((64*\t)/x)}}

\begin{tikzpicture}[scale=1.0,font=\small]
\begin{axis}[name=main plot,scale only axis, 
	     width=3.50cm,  height=3.0cm,
	     ylabel={$P\ [\rm atm]$}, 
	     y label style={at={(axis description cs:-0.25,.5)},rotate=0,anchor=south},
	     xlabel={$V_{conf}\ [\rm nm^3]$}, 
	     ymin=0, 	ymax=2000, ytick distance=500,
	     xmin=1.85, xmax=1.93, xtick distance=0.02,
             axis line style={black, line width=0.25mm},
	     every tick label/.append style={black},
	     every axis label/.append style ={black},
	     every tick/.style={line width=0.30mm},
    	     xticklabel=\pgfkeys{/pgf/number format/.cd,fixed,precision=2,zerofill}\pgfmathprintnumber{\tick},
    	     name=ax1,
	     ]
	   
\def\a{-0.9787}
\def\b{6.55e+10}
\def\c{6.74e+09}
\def\d{-23.2678}
\def\t{300}
\def\pref{6070.3}
\addplot[
    domain=1.84:1.925,
    samples=20,
    color=azul,
    name path=f,
    line width=1.2pt
    ]
    {pVinet-\pref};

\def\a{-0.9605}
\def\b{6.55e+10}
\def\c{6.74e+09}
\def\d{-23.2274}
\def\t{315}
\def\pref{6070.3}
\addplot[
    domain=1.84:1.925,
    samples=20,
    color=verde,
    name path=f,
    line width=1.2pt
    ]
    {pVinet-\pref};
    
\def\a{-0.9412}
\def\b{6.50e+10}
\def\c{6.74e+09}
\def\d{-23.2123}
\def\t{330}
\def\pref{6070.3}
\addplot[
    domain=1.84:1.925,
    samples=20,
    color=rojo,
    name path=f,
    line width=1.2pt
    ]
    {pVinet-\pref};

  \coordinate (c1) at (axis cs:1.900,0);
  \coordinate (c2) at (axis cs:1.900,500);
  \draw[black, line width = 0.2mm, fill opacity=0.05, fill=magenta] (c1) rectangle (axis cs:1.925,500);
\end{axis}

\begin{axis}[
   name=ax2, width=4.75cm,  height=4.0cm,  
   scaled ticks=false,
   xmin=1.905,xmax=1.925,
   ymin=0, ymax=100,
   ylabel={$P=\text{c}$} , 
   xlabel={$V_{\text{conf}}~[\text{nm}^{3}]$}, 
   x label style={at={(axis description cs:-0.2,-0.1)},anchor=north},
   y label style={font=\scriptsize, at={(axis description cs:-0.12,.12)},rotate=-90,anchor=south},   
   xtick=\empty, ytick=\empty, 
   xlabel=\empty,
   axis line style={black, line width=0.25mm},
   at={($(ax1.south east)+(0.75cm,0.5cm)$)}, axis background/.style={fill=magenta, fill opacity=0.05},
   clip=false
 ]
\draw[dashed, line width=0.75pt] (1.905,20) -- (1.925,20);

\def\a{-0.9787}
\def\b{6.55e+10}
\def\c{6.74e+09}
\def\d{-23.2678}
\def\t{300}
\def\pref{6070.3}
\addplot[
    domain=1.905:1.9074,
    samples=2,
    color=azul,
    name path=f2,
    line width=1.2pt
    ]
    {pVinet-\pref};
\node[fill=none,rotate=-76,line width=1.2pt] at (1.9075, 29) {\scriptsize\textcolor{azul}{$T = 300\ \rm K$}};
\draw[gray, dashed, line width=0.75pt] (1.9065,0)--(1.9065,20);
\node[circle, draw=azul, fill=azul!25, minimum size=0.5pt, line width=0.25mm,scale=0.45] at (1.9065,20) {};
\node[circle,fill=black,inner sep=0pt,minimum size=1.25pt,scale=0.6,label=below:{\tiny\textcolor{azul}{$\boldsymbol{V_{conf}^{(1)}}$}}] at (1.9065,0) {};
\def\a{-0.9605}
\def\b{6.55e+10}
\def\c{6.74e+09}
\def\d{-23.2274}
\def\t{315}
\def\pref{6070.3}
\addplot[
    domain=1.915:1.91905,
    samples=2,
    color=verde,
    name path=f,
    line width=1.2pt
    ]
    {pVinet-\pref};
\node[fill=none,rotate=-76] at (1.9182, 50) {\scriptsize\textcolor{verde}{$T = 315\ \rm K$}};
\draw[gray, dashed,line width=0.75pt] (1.91825,0)--(1.91825,20);
\node[circle, draw=verde, fill=verde!25, minimum size=0.5pt, line width=0.25mm,scale=0.45] at (1.91825,20) {};
\node[circle,fill=black,inner sep=0pt,minimum size=1.25pt,scale=0.6,label=below:{\tiny\textcolor{verde}{$\boldsymbol{V_{conf}^{(2)}}$}}] at (1.91825,0) {};

\def\a{-0.9412}
\def\b{6.50e+10}
\def\c{6.74e+09}
\def\d{-23.2123}
\def\t{330}
\def\pref{6070.3}
\addplot[
    domain=1.9197:1.9235,
    samples=2,
    color=rojo,
    name path=f,
    line width=1.2pt
    ]
    {pVinet-\pref}; 
\node[fill=none,rotate=-79,line width=1.2pt] at (1.922, 71) {\scriptsize\textcolor{rojo}{$T = 330\ \rm K$}};
\draw[gray, dashed, line width=0.75pt] (1.9227,0)--(1.9227,20);
\node[circle, draw=rojo, fill=rojo!25, minimum size=1.25pt, line width=0.25mm,scale=0.45] at (1.9227,20) {};
\node[circle,fill=black,inner sep=0pt,minimum size=1.25pt,scale=0.6,label=below:{\tiny\textcolor{rojo}{$\boldsymbol{V_{conf}^{(3)}}$}}] at (1.9227,0) {};
\end{axis}
\draw [dashed] (c1) -- (ax2.south west);
\draw [dashed] (c2) -- (ax2.north west);
\end{tikzpicture}  
\caption {The $P$ vs $V$ curves are shown for the isotherms 300, 315, 330 K. The inset is a zoom projection of $P$ and $V$. The dashed line illustrates a constant 
value of the pressure $P$.} \label{fig.04}
\end{figure}
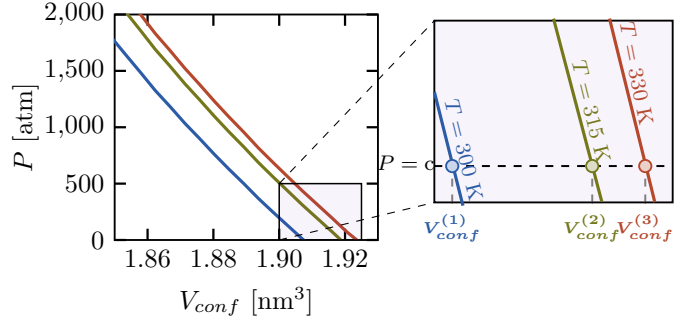

\medskip

\noindent The pressure values 1, 50 and 100 atm are in the range resulting from the compression process. We determine the confining volume and internal energy on the 
different isotherms to establish the enthalpy values ({\color{MyColor2} Fig. \ref{fig.04}}):

\vspace{-0.5cm}
\begin{equation} \label{conprepro.1}
\begin{tabular} {c}
$H_1 (T_1= 300\, {\rm K})= U_1 + PV_{conf}^{(1)}$\\\\[-0.3cm]
$H_2 (T_2= 315\, {\rm K})= U_2 + PV_{conf}^{(2)}$\\\\[-0.3cm]
$H_3 (T_3= 330\, {\rm K})= U_3 + PV_{conf}^{(3)}$
\end{tabular}
\end{equation}
\vspace{-0.3cm}

\noindent The enthalpy values are reported in {\color{MyColor1} Table \ref{table.04}}. The relation of the enthalpy to the temperature at constant pressure is found to 
be a linear expression, $H(T)= AT + B$, in the range of investigated temperatures. The values 20.84, 20.85, and 20.85 in units of cal/(mol K) are obtained for the 
pressures 1 atm, 50 atm, and 100 atm, respectively. The experimental values are 17.99 cal/(mol K) at 1 atm, and 17.89 cal/(mol K) at 100 atm \cite{Kell1975}. There are 
reported values of $C_P$ in the literature using different model potentials for the water interactions. The values of those studies show larger deviations from the 
experimental values \cite{Wu2004, Vega2005, Vega2010}.

\medskip

\noindent With a pressure of 1 atm, we have $C_V$= 4.6485 J/(g K) and $C_P$= 4.8400 J/(mol K), and $C_P > C_V$, which is consistent with the thermodynamics relations.

\section {$\beta$ and $\kappa$ volumetric response coefficients} \label{volrescof}

\begin{figure} [!b]
\centering
\input{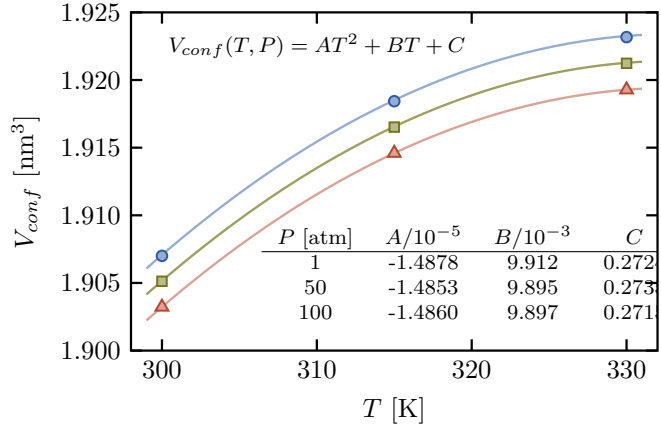}

\pgfmathdeclarefunction{line}{0}{\pgfmathparse{\a*x^2 + \b*x + \c}}
\begin{tikzpicture}[scale=1.0,font=\small]
\begin{axis}[name=main plot,scale only axis, 
	     width=7.0cm,  height=4.5cm,
	     xlabel={$T~[\rm K]$}, ylabel={$V_{conf}~[\rm nm^{3}]$ }, 
	     ymin=1.90, ymax=1.925, ytick distance=0.005,
	     xmin=298, 	xmax=332, 	xtick distance=10,
	     line width=0.4pt, 
	     axis line style={black, line width=0.30mm},
	     every tick/.style={line width=0.30mm},
	     every tick label/.append style={black},
	     every axis label/.append style ={black},
	     yticklabel=\pgfkeys{/pgf/number format/.cd,fixed,precision=3,zerofill}\pgfmathprintnumber{\tick},
	     ]   
\node[fill=none,rotate=0] at (310,1.9225) {\scriptsize\textcolor{black}{$V_{conf}(T,P) = AT^{2} + BT + C$}};
\def\a{-0.000014877}
\def\b{0.009912}
\def\c{0.2724}
\addplot[
    domain=299:331,
    samples=200,
    color=azul!50,
    name path=f,
    line width=0.90pt
    ]
    {line};
\addplot+[only marks, mark=*,  
          azul, 
          mark options={fill=azul!50, solid}, 
          mark size=2.0pt, line width=0.75pt
         ] table[x index=0,y index=1] {vconf-temp.dat};
\node[fill=none,rotate=28.5] at (305,2.315) {\textcolor{azul}{$P = 1~\text{atm}$}};
\def\a{-0.000014853}
\def\b{0.0098947}
\def\c{0.273512}
\addplot[
    domain=299:331,
    samples=200,
    color=verde!70,
    name path=f,
    line width=0.90pt
    ]
    {line};
\addplot+[only marks,  mark=square*,  
          verde, 
          mark options={fill=verde!50, solid}, 
          mark size=1.8pt, line width=0.75pt
         ] table[x index=0,y index=2] {vconf-temp.dat};
\node[fill=none,rotate=28.5] at (315,2.3165) {\textcolor{verde}{$P = 50~\text{atm}$}};
\def\a{-0.00001486}
\def\b{0.0098969667}
\def\c{0.271535}
\addplot[
    domain=299:331,
    samples=200,
    color=rojo!50,
    name path=f,
    line width=0.90
    ]
    {line};
\addplot+[only marks,mark=triangle*,  
          rojo, 
          mark options={fill=rojo!50, solid}, 
          mark size=2.8pt, line width=0.75pt
         ] table[x index=0,y index=3] {vconf-temp.dat} ;
\node[fill=none,rotate=27.5] at (325,2.315) {\textcolor{rojo}{$P = 100~\text{atm}$}};

\node at (320,1.9055) {\scriptsize%
\begin{tabular}{cccc}
 $P\ [\rm atm]$ & $A/10^{-5}$ & $B/10^{-3}$ & $C$ \\   \cline{1-4} 
1   & -1.4878 & 9.912 & 0.2724 \\
50  & -1.4853 & 9.895 & 0.2735 \\
100 & -1.4860 & 9.897 & 0.2715 \\
\end{tabular}
};
\end{axis}
\end{tikzpicture}  
\caption {The confinement volume $V_{conf}$ as a function of the temperature $T$ is shown. The interpolating curves match the numerical data. The fitting coefficients 
$A\, [{\rm nm}^3/{\rm K}^2]$, $B\, [{\rm nm}^3/{\rm K}]$, and $C\, [{\rm nm}^3]$ are shown for different pressure values.} \label{fig.05}
\end{figure}

\noindent The $\beta$ response factor represents the volume change of the system with respect to the temperature, keeping the pressure constant. {\color{MyColor1} 
Table \ref{table.04}} lists the required observables to determine $\beta$ with fixed pressure. The data  have a non-linear trend ({\color{MyColor2} Fig. \ref{fig.05}}). 
A quadratic expression fits the data in the analyzed temperature range. Thereby, $\partial V_{conf}/ \partial T= 2AT + B$, and the thermal response coefficient 
is $\beta (V,T)= (2AT + B)/ V_{conf} (T, P=$ constant). Under the thermodynamic conditions $(P,V_{conf},T)= (1\, {\rm atm},\ 1.9070\, {\rm nm}^3,\ 300\, {\rm K})$, the 
water-bulk value of $\beta$ is $0.5166 \times 10^{-3}\ {\rm K}^{-1}$. The experimental value is $0.2728 \times 10^{-3}\ {\rm K}^{-1}$, which is of the same order of 
magnitude for the small water cluster (recall that it is the worst-scenario case).

\medskip

\noindent In order to determine the isothermal compressibility coefficient $\kappa$, Eq. (\ref{mechpre.2}) is employed to obtain $(\partial P/\partial V)_T$ analytically. 
By using the identity $(\partial V/ \partial P) _T = 1/ (\partial P/ \partial V)_T$ \cite{Callen1985}, and inserting it in the expression of $\kappa$, we obtain:

\vspace{-0.5cm}
\begin{equation} \label{volrescof.1}
\displaystyle 
\kappa(V,T)= \bigg[e^{\gamma V^{1/3}} \left( \frac {b \gamma^2} {9 V^{1/3}} + \frac {c \gamma^2} 9 -\frac {2b \gamma} {9V^{2/3}}
-\frac {2c} {9 V^{2/3}} \right) + \frac {N k_B T_{eq}} V\bigg]^{-1}
\end{equation}
\vspace{-0.25cm}

\noindent The parameters $b$, $c$, and $\gamma$ are known from {\color{MyColor1} Table \ref{table.03}}. By using $\kappa$ of Eq. (\ref{volrescof.1}), and providing values 
of the temperature and confining volume, the factor $\kappa$ is evaluated. When $T= 300\, {\rm K}$ and $V_{conf}= 1.907\, {\rm nm}^3$, we have 
$\kappa= 0.1993\, {\rm GPa}^{-1}$. The experimental value is $0.4511\, {\rm GPa}^{-1}$. We emphasize that the system of this work is a worst-scenario case, because it is 
sufficiently small with respect to the size of a bulk system.

\section {Entropy and thermodynamic potentials} \label{enathpot}

\begin{figure} [!t]
\centering
\begin{tikzpicture}[scale=1.0,font=\small,
                    forward/.style={thick,->,shorten >=2pt,>=stealth}]
\begin{axis}[name=main plot,scale only axis, 
	     width=7.0cm,  height=5.5cm,
	     ylabel={$P\ [\rm atm]$}, 
	     xlabel={$V_{conf}\ [\rm nm^{3}]$}, 
	     ymin=-40, 	ymax=140, ytick distance=25.0,
	     ytick={0,25,50,75,100},
	     xmin=1.90, 	xmax=1.925, 	xtick distance=0.005,
	     line width=0.4pt, 
	     axis line style={black, line width=0.30mm},
	     every tick/.style={line width=0.30mm},
	     every tick label/.append style={black},
	     every axis label/.append style ={black},
    	     xticklabel=\pgfkeys{/pgf/number format/.cd,fixed,precision=3,zerofill}\pgfmathprintnumber{\tick},
    	     name=ax1,
	     ]
\path[name path=axis] (axis cs:2.27,0) -- (axis cs:2.3512,0);
\draw[forward,line width=1.2pt] (1.9070,1.0) -- (1.923170,1.0);
\draw[forward,line width=1.2pt,rojo] (1.923170,1.0) -- (1.919280,100);
\draw[forward,line width=1.2pt] (1.919280,100) -- (1.903225,100);
\draw[forward,line width=1.2pt,azul] (1.903225,100) -- (1.9070,1.0);

\node  [single arrow, 
        top color=rojo, bottom color=white, 
        minimum height=1.0cm, single arrow tip angle=100,  single arrow head indent=0.05cm,
        rotate=90] at (1.917,0) {};
\node[fill=none,rotate=0]   at (1.917,-25)  {\footnotesize\textcolor{rojo}{$\boldsymbol{Q_{P_{1}}}$}};
\node  [single arrow, 
        top color=azul, bottom color=white, 
        minimum height=1.0cm, single arrow tip angle=100,  single arrow head indent=0.05cm,
        rotate=90] at (1.907,100) {};
\node[fill=none,rotate=0]   at (1.907,127)  {\footnotesize\textcolor{azul}{$\boldsymbol{ Q_{P_{2}} }$}};

\node  [single arrow, 
        left color=black!70, right color=white, 
        minimum height=0.8cm, single arrow tip angle=100,  single arrow head indent=0.05cm,
        rotate=180] at (1.9202,75) {};
\node[fill=none,rotate=0]   at (1.923,75)  {\footnotesize\textcolor{black}{$W^{{\rm in}}$}};

\node  [single arrow, 
        left color=black!70, right color=white, 
        minimum height=0.8cm, single arrow tip angle=100,  single arrow head indent=0.05cm,
        rotate=180] at (1.906,20) {};
\node[fill=none,rotate=0]   at (1.9025,20)  {\footnotesize\textcolor{black}{$W^{{\rm out}}$}};

\node[circle, draw=black, fill=white, minimum size=0.5pt, line width=1.2pt, scale=0.5] at (1.9070,1) {};
\node[circle, draw=black, fill=white, minimum size=0.5pt, line width=1.2pt, scale=0.5] at (1.9032,100) {};   

\node[circle, draw=black, fill=white, minimum size=0.5pt, line width=1.2pt, scale=0.5] at (1.923170,1) {};
\node[circle, draw=black, fill=white, minimum size=0.5pt, line width=1.2pt, scale=0.5] at (1.919280,100) {};    
\node[fill=none,rotate=0] at (1.9063, -5) {\footnotesize \textbf{1}};
\node[fill=none,rotate=0] at (1.924, -5) {\footnotesize \textbf{2}};
\node[fill=none,rotate=0] at (1.92,105) {\footnotesize \textbf{3}};
\node[fill=none,rotate=0] at (1.9023,105) {\footnotesize \textbf{4}};
\node[fill=none,rotate=0]   at (1.9115,8)    {\footnotesize $P_{1}=1\ \rm atm$};
\node[fill=none,rotate=-70] at (1.9209,35)   {\footnotesize \textcolor{rojo}{$\boldsymbol{T_{2} = 330\ {\rm K}}$}};
\node[fill=none,rotate=0]   at (1.915,108) {\footnotesize$P_{2} = 100\ \rm atm$};
\node[fill=none,rotate=-72] at (1.9035,68)   {\footnotesize \textcolor{azul}{$\boldsymbol{T_{1} = 300\ {\rm K}}$}};
\end{axis}
\end{tikzpicture}  
\caption {$PV$ diagram of a proposed thermodynamic cycle. The arrows indicate the exchange of work and heat of the confined particle system with the heat reservoir.} 
\label{fig.06}
\end{figure}
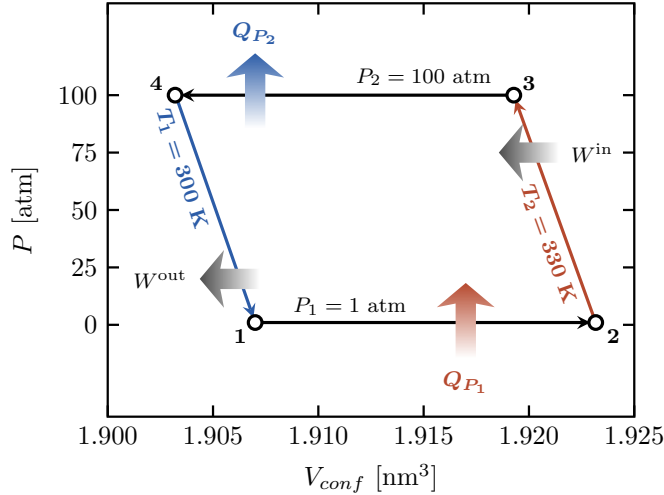

\noindent The following goal is to evaluate the changes of the internal energy $U$, enthalpy $H$, and entropy $S$ in a cyclic thermodynamic process, which involves 
changes of $P$, $V_{conf}$ and $T$. The thermodynamic cycle consists of four stages illustrated in the $PV$ diagram of {\color{MyColor2} Fig. \ref{fig.06}}. The 
isothermal curves are essentially straight lines and $P= AV_{conf} + B$. The values of the independent variables and energy functions are given in {\color{MyColor1} 
Table \ref{table.05}}. 

\medskip

\noindent The values of $U$ and $H$ are available from the previous calculations, and the changes of $U$ and $H$ in the thermodynamic cycle are directly computed, 
$\Delta U= U_f -U_i$ and $\Delta H= H_f -H_i$. For constant pressure processes, the enthalpy changes $\Delta H_{12}$ and $\Delta H_{34}$ correspond to the heat 
$Q_{P_1}$ absorbed by the system in process $1 \rightarrow 2$, and to the heat $Q_{P_2}$ released by the system in process $3 \rightarrow 4$, respectively. According to 
classical thermodynamics, it is possible to estimate the entropy changes $\Delta S$ in terms of the mechanical response factors. {\color{MyColor1} Table \ref{table.06}} 
shows the changes of $U$, $H$, and $S$ in the four processes of the cycle. The total change of internal energy in the closed thermodynamic cycle is basically zero, and 
similarly in the case of $\Delta H$ and $\Delta S$, a characteristic of \emph{state functions} in thermodynamics. The estimates using the NIST information (Appendix B) 
are also provided there.

\begin{table}[b]
\tbl {Physical observables of the thermodynamic cycle schematized in {\color{MyColor2} Fig. \ref{fig.06}}.$^\ast$} 
{\begin {tabular} {lcccccccc} \toprule
macroscopy && \multicolumn {7} {c} {thermodynamic states}\\
\cline {3-9}
\multicolumn {1} {c} {property} && 1 && 2 && 3 && 4\\
\cline {1-1} \cline {3-3} \cline {5-5} \cline {7-7} \cline {9-9}
$P$ [atm]           &&   1       &&  1       &&  100     &&  100\\
$T$ [K]             &&   300     &&  330     &&  330     &&  300\\
$V_{conf}$ [nm$^3$] &&   1.907   &&  1.923   &&  1.919   &&  1.903\\
$E_p$ [au]          &&  -0.9570  && -0.9199  && -0.9195  && -0.9565\\
$U$ [au]            &&  -0.6854  && -0.6216  && -0.6212  && -0.6850\\
$H$ [au]            &&  -0.6853  && -0.6216  && -0.6167  && -0.6805\\ \bottomrule
\end{tabular}}
\tabnote{$^\ast$Up to four digits after the decimal point are presented with the purpose of observing differences in the energies.
}
\label{table.05}
\end{table}

\begin{table}[!t]
\tbl {Changes of $U$, $H$ and $S$ in the thermodynamic cycle of {\color{MyColor2} Fig. \ref{fig.06}}.$^\ast$} 
{\begin {tabular} {ccccccc} \toprule
process && $\Delta U$ [kJ/mol] && $\Delta H$ [kJ/mol] && $\Delta S$ [J/(mol K)]\\
\cline{1-1} \cline{3-3} \cline{5-5} \cline{7-7}
$1 \rightarrow 2$ &&  2.616 (2.259)       &&  2.616 (2.259)       &&  8.310 (7.175)\\
$2 \rightarrow 3$ && \cellcolor{gray!20} 0.017 (0.335)   && \cellcolor{gray!20} 0.198 (0.153)   && \cellcolor{gray!20} -0.009 (-0.092)\\
$3 \rightarrow 4$ && -2.626 (-2.248)      && -2.618 (-2.246)      && -8.314 (-7.134)\\
$4 \rightarrow 1$ && \cellcolor{gray!20} -0.017 (-0.346) && \cellcolor{gray!20} -0.196 (-0.166) && \cellcolor{gray!20} 0.093 (0.051)\\
total             &&  0.010 (0.000)       &&  0.00  (0.000)       &&  0.080 (0.000)\\ \bottomrule
\end{tabular}}
\tabnote{$^\ast$The values in parenthesis are deduced from the NIST thermodynamic data, {\color{MyColor1} Table \ref{table.08}}. From such data, the internal energy change 
is calculated in the form $\Delta U= \Delta H -\Delta (PV)$. For example for the process $1 \rightarrow 2$, we have $U_2 -U_1= (H_2 -H_1) -(P_2 V_2 -P_1 V_1)$. Refer 
to Sect. \ref{enathpot} for the discussion on the values in a gray background.
}
\label{table.06}
\end{table}

\begin{table}[b]
\tbl {Free energy changes and pressure along the isothermal curves of the thermodynamic cycle of {\color{MyColor2} Fig. \ref{fig.06}}.$^\ast$} 
{\begin {tabular} {clcccc} \toprule
&&& $2 \rightarrow 3$ && $4 \rightarrow 1$\\
\multicolumn {2} {c} {free energies [kJ/mol]} && (330 K) && (300 K)\\
\cline{1-2} \cline{4-4} \cline{6-6}
$\Delta F$\\
& NIST && 0.3654 && -0.3613\\
& $\displaystyle \Delta F = -\int_i^f PdV$ && 0.1870 && -0.1815\\

$\Delta G$\\
& NIST && 0.1834 && -0.1813\\
& $\displaystyle \Delta G = \int_i^f VdP$ && 0.1814 && -0.1798\\

$P= AV_{conf} + B$\\
& A $[{\rm atm}/{\rm nm}^3]$ && $-2.6225 \times 10^4$ && $-2.5450 \times 10^4$\\
& B $[{\rm atm}]$            &&  $5.0013 \times 10^4$ &&  $4.8946 \times 10^4$\\ \bottomrule
\end{tabular}}
\tabnote{$^\ast$The expression of $P$ reproduces the pressure in going from state 2 to state 3, and from state 4 to state 1 of the thermodynamic cycle. The values of $A$ 
and $B$ are computed from data of Table \ref{table.04}.
}
\label{table.07}
\end{table}

\medskip

\noindent As it can be observed from the results of {\color{MyColor1} Table \ref{table.06}}, the small energies (below 1 kJ/mol and denoted with a gray background) show 
differences with the values obtained from the NIST information. Such deviations are associated to two main factors: $(i)$ the TIP3P force field of water is known to have 
a limited prediction of the energies, specially when the particle system is subjected to different temperatures \cite{Carlos2011, Saeed2016}, and $(ii)$ in the process 
of doing energy fittings (with the purpose of obtaining analytical expressions of certain energies), small deviations between the predicted energies from the analytical 
expressions and the numerical energies occur. In consequence, the accumulation of such deviations may be the reason to obtain energy differences smaller than 1 kJ/mol. 
Note that 1 kJ/mol is lower than the chemical accuracy expected, for example, from density functional theory. In this regard, the small energies of 
{\color{MyColor1} Table \ref{table.06}} should be interpreted with care, and the deviations should not be associated to the formulation of this work because the numerical 
energies may be improved under the use of better force fields or accurate methods of electronic structure theory.

\medskip

\noindent The changes of free energies $\Delta F$ and $\Delta G$ are obtained after integrations of $P$ and $V$, respectively. The integration of $P$ is performed using 
the analytic form of the pressure in terms of the volume:

\vspace{-0.65cm}
\begin{equation} \label{enathpot.1}
\begin{tabular} {c}
$\displaystyle  \Delta F= \Delta E_p -N k_B T_{mec} \ln \left( \frac {V_2} {V_1} \right)$
\end{tabular}
\end{equation}
\vspace{-0.25cm}

\noindent In the case of the Gibbs energy, we require the inverse function $V= V(P,T)$, which is generated from the data of {\color {MyColor1} Table \ref{table.04}}. 
Instead of using the state functions $P= P(V,T)$ and $V= V(P,T)$, the free energy changes may be alternatively computed from $\Delta U$, $\Delta H$, and $\Delta S$, and 
the listed values of {\color{MyColor1} Table \ref{table.06}}. The values of $\Delta F$ and $\Delta G$ for the isothermal processes 
$2 \rightarrow 3$ and $4 \rightarrow 1$ of the thermodynamic cycle are shown in {\color{MyColor1} Table \ref{table.07}} and compared with those generated from the NIST 
data.

\medskip

\noindent It is important to mention that the formulation of this work gives foundation to previous molecular models where: $(i)$ the results on highly pressurized 
hydrogen clusters have shown excellent agreement with the experimental results using diamond anvil cells \cite{Santamaria2005}, and $(ii)$ a single electron 
(an elementary particle) was confined in a spherical cage of He atoms, showing the localized and delocalized states of the electron by simply changing the cage radius, 
thus exhibiting the wave-particle duality of the electron by changing a single parameter in the simulation \cite{Santamaria2019}.

\section {Conclusions} \label{conclusions}

\noindent The formulation presented in this work combines a Lagrangian description of the confined atoms in interaction with the container atoms, and a Langevian 
description of the container atoms in interaction with the heat bath. The combination of such levels of theory takes us from an original set of expressions, where all of 
them are deterministic, to a subset of equations where the deterministic character is maintained for the confined atoms, and another subset of equations where the initial 
expressions have been transformed onto stochastic equations of motion for the atoms forming the container. All the equations of motion are coupled together, and have 
universal character as they are not restricted to a given number of atoms, type of atom, or specif form of atomic arrangement. In this quest the molecular nature of 
thermodynamics is exhibited, and confirmed by reproducing the known features of heat, the first law of thermodynamics, and allowing to introduce in mechanical form the 
energetic and mechanical response coefficients, pressure, amount of heat transfer, entropy, and free energies of the system of confined particles. In this respect, 
the formulation is different to other approaches reported in the literature. The numerical results on a thermodynamic prototype, intentionally designed to be a 
worst-scenario case due to its small size, use of moderate forcefields, and influence of the border effects, are relatively close to the experimental measurements on 
water bulk. The present formulation gives the possibility of obtaining thermodynamic properties of systems of nanometric sizes from molecular simulations by simply 
imitating the ordinary measurement steps performed in the laboratory. The formulation may be also used to investigate systems far from thermodynamic equilibrium due to 
the general nature of molecular dynamics. Preliminary results on the application of this formulation in fluid dynamics has shown excellent results.

\section {Acknowledgments} \label{acknowledgments}

\noindent RS and EHH thank DGTIC-UNAM for access to the Miztli-UNAM computer, and express gratitude to Carlos E. Lopez Nataren and the staff of IF-UNAM for helping with 
the computations. RS and EHH acknowledge PAPIIT-DGAPA (Project Num. IN-111-918) for financial support.
\section {Appendix A: model interaction potentials} \label{modintpot}

\noindent The model potentials that simulate the interactions of the confined particles and the container particles are described in this section. The model potentials 
are contributions to the energy $E_{xs}$, Eq. (\ref{tlagpro.3}).

\subsection {Potential energy of the container atoms} \label{tpeotca}

\noindent The interactions among the atoms are considered of harmonic type. Consider atom 1, whose bonded neighbors are atoms 2, 3, and 4, with positions 
${\bf x}_2$, ${\bf x}_3$, and ${\bf x}_4$. The antipodal atom to atom 1 is atom 5, with position ${\bf x}_5$. The potential energy of atom 1 in  interaction with the 
other atoms is:

\vspace{-0.5cm}
\begin{equation} \label{tpeotca.1}
\begin{tabular} {c}
$\displaystyle  U^{str} ({\bf x}_1)= \frac 1 2\, k_{bond} \left[ ( |{\bf x}_2 -{\bf x}_1|- r_0)^2 +\ ( |{\bf x}_3 -{\bf x}_1| -r_0)^2 + ( |{\bf x}_4 -{\bf x}_1| -r_0)^2 \right] $\\\\[-0.25cm]


$\displaystyle  +\ \frac 1 2\, k_{dist} ( |{\bf x}_5 -{\bf x}_1|- R_0)^2$
\end{tabular}
\end{equation}
\vspace{-0.2cm}

\noindent The total potential energy of the container is obtained by addition of potentials of the above type, 
$U_x^{str}= U^{str} ({\bf x}_1) + U^{str} ({\bf x}_2) + U^{str} ({\bf x}_3) + \cdots$. The container radius was reduced in different simulations, resulting in changes of 
the interatomic equilibrium distances $r_0$ and $R_0$, but maintaining the same spring stiffness $k_{bond}$ and $k_{dist}$ of the bonded atoms and the antipodal atom, 
respectively. The equilibrium distances used in the simulations for $r_0$ are 1.73, 1.72, 1.69, 1.65, and 1.62 in \AA, while for $R_0$ they are 
17.83, 17.64, 17.30, 16.95, 16.60 in \AA. The magnitude of the spring stiffness between both the bonded atoms and the antipode atoms is 1066.737 and 43.926 in 
kcal/(mol\, \AA$^2$), respectively.

\subsection {Potential energy of the container-confined atoms} \label{tpecaica}

\noindent The interactions of the container atoms and the confined atoms are simulated with the van der Waals ($vdw$) potential. Consider the container atom 1 with 
position ${\bf x}_1$ as a reference atom, and an arbitrary water molecule with atoms O, H$_2$, H$_3$ and positions ${\bf s}_1$, ${\bf s}_2$, and ${\bf s}_3$. The 
interaction energy of the reference atom with that water molecule is:

\vspace{-0.7cm}
\begin{equation} \label{tpecaica.1}
\begin{tabular} {c}
$\displaystyle  U^{vdw} ({\bf x}_1, {\bf s}_1, {\bf s}_2, {\bf s}_3)= 4 \varepsilon_{{\rm CO}} \left[ \left( \frac {\sigma_{{\rm CO}}}   {| {\bf x}_1 -{\bf s}_1| } \right)^{12} -\left( \frac {\sigma_{{\rm CO}}} {| {\bf x}_1 -{\bf s}_1 |} \right)^6 \right]$\\\\[-0.3cm]

$\displaystyle +\ 4 \varepsilon_{{\rm CH}} \left[ \left( \frac {\sigma_{{\rm CH}_1}} {| {\bf x}_1 -{\bf s}_2 |} \right)^{12} -\left(\frac {\sigma_{{\rm CH}_1}} {|{\bf x}_1 -{\bf s}_2| }\right)^6\right]$\\\\[-0.3cm]

$\displaystyle +\ 4 \varepsilon_{{\rm CH}} \left[ \left( \frac {\sigma_{{\rm CH}_2}} {| {\bf x}_1 -{\bf s}_3 |} \right)^{12} -\left(\frac {\sigma_{{\rm CH}_2}} {| {\bf x}_1 -{\bf s}_3 |} \right)^6 \right]$
\end{tabular}
\end{equation}
\vspace{-0.3cm}

\noindent The $vdw$ potential contains two parameters per atomic pair, they are $\varepsilon_{\rm CO}$ and $\sigma_{\rm CO}$ for the $\rm C \cdots O$ interaction, and 
whose values in that order are 0.1039 kcal/mol and 3.372 \AA, and $\varepsilon_{\rm CH}$ and $\sigma_{\rm CH}$ for the $\rm C \cdots H$ interaction with values of 
0.0256 kcal/mol and 2.640 \AA. \cite{Yanbin2013} The interactions of the container atom 1 with all the confined water molecules are obtained by addition of the 
potentials of the above type, $U^{vdw} ({\bf x}_1, \{ {\bf s}_k \})= 4 \varepsilon \sum_k [(\sigma/ |{\bf x}_1 -{\bf s}_k| )^{12} -(\sigma/ |{\bf x}_1 -{\bf s}_k| )^6]$. 
The total interaction energy of the container with the confined particles is obtained by summing the interactions over all the container atoms, 
$U_{xs}^{vdw}= U^{vdw} ({\bf x}_1, \{ {\bf s}_k \}) + U^{vdw} ({\bf x}_2, \{ {\bf s}_k \}) + \cdots$.

\subsection {Potential energy of the confined atoms} \label{tpeotcona}

\noindent The potential energy of the confined particles is due to the intramolecular energy and the intermolecular energy. The intramolecular energy is simulated by 
harmonic potentials. Consider a water molecule whose atoms have positions ${\bf s}_1$, ${\bf s}_2$, and ${\bf s}_3$. The intramolecular energy of that molecule is:

\vspace{-0.5cm}
\begin{equation}\label{intrapotene.1}
\begin{aligned} 
\displaystyle 
U^{str} ({\bf s}_1, {\bf s}_2, {\bf s}_3) =& \frac 1 2\, k_{{\rm OH}} \left(| {\bf s}_1 -{\bf s}_2| -r_{{\rm OH}_1}\right)^2 +\ \frac 1 2\, k_{{\rm OH}} (|{\bf s}_1 -{\bf s}_3| -r_{{\rm OH}_2})^2 \\ 
& + \ \frac 1 2\, k_{{\rm HOH}} [\theta ({\bf s}_1, {\bf s}_2, {\bf s}_3) -\theta_{{\rm HOH}}]^2
\end{aligned} 
\end{equation}
\vspace{-0.3cm}

\noindent The first two terms include the harmonic stretching of bonds ${\rm OH}_1$ and ${\rm OH}_2$, and the last term describes the harmonic stretching of the angle 
HOH. The equilibrium distances are $r_{{\rm OH}_1}$ and $r_{{\rm OH}_2}$, with $r_{{\rm OH}_1}= r_{{\rm OH}_2}=$ 0.9572 \AA. The $\theta_{{\rm HOH}}$ equilibrium angle 
is 104.52 Deg. The respective force constants are $k_{{\rm OH}_1}$, $k_{{\rm OH}_2}$, with $k_{{\rm OH}_1}= k_{{\rm OH}_2}=$ 1207.0720 kcal/(mol \AA$^2$), and 
$k_{{\rm HOH}}=$ 150.0998 kcal/(mol Rad$^2)$. The values of these parameters correspond to the TIP3P water model \cite{Klein1983}. For the $N$ confined water molecules, 
we have the total intramolecular potential energy $U_s^{str}= U^{str} ({\bf s}_1, {\bf s}_2, {\bf s}_3) + U^{str} ({\bf s}_4, {\bf s}_5, {\bf s}_6) + \cdots$

\medskip

\noindent The intermolecular potential has electrostatic and van der Waals contributions. The electrostatic interaction is due to nonbonded charged atoms:

\vspace{-0.5cm}
\begin{equation} \label{interpoten.1}
\begin{tabular} {c}
$\displaystyle U_s^{el}= \frac 1 2 \sum_{i, j} k_c\, \frac {q_i q_j} {| {\bf s}_i -{\bf s}_j |} \hspace{1.0cm} ; \hspace{1.0cm} i \neq j$
\end{tabular}
\end{equation}
\vspace{-0.3cm}

\noindent The indexes $i$ and $j$ run over atoms of different water molecules. The potential is the TIP3P water coulombic potential with parameters 
$q_{\rm H}$, $q_{\rm O}$ and $k_{\rm c}$, and values of $+0.417\, e$, $-0.834\, e$, and  332.0595 (kcal/mol) (\AA/ e$^2$). The other contribution is due to the 
interaction of the electric dipoles of the water molecules. It is simulated using a TIP3P-water $vdw$ potential. The parameters for the $\rm O \cdots O$ interaction 
are $\varepsilon_{\rm OO}$ and $\sigma_{\rm OO}$, with values in that order of 0.1521 kcal/mol and 3.1507 \AA \cite{Klein1983}.

\vspace{-0.5cm}
\begin{equation} \label{interpoten.2}
\begin{tabular} {c}
$\displaystyle U_s^{vdw}= 4 \varepsilon_{{\rm OO}} \sum_{i, j} \left[ \left( \frac {\sigma_{{\rm OO}}} {|{\bf s}_i -{\bf s}_j|} \right)^{12} -\left( \frac {\sigma_{{\rm OO}}} {| {\bf s}_i -{\bf s}_j |} \right)^6 \right]$
\end{tabular}
\end{equation}
\vspace{-0.2cm}

\section {Appendix B: thermodynamic data from NIST} \label{tpftnd}

\noindent Each of the four thermodynamic states in the cycle of Fig. \ref{fig.06} is defined by the temperature and pressure. {\color{MyColor1} Table \ref{table.08}} 
gives $V$, $H$, and $S$ in terms of the temperatures and pressures.

\begin{table} [t]
\tbl {Thermodynamic properties of bulk water from the NIST data.$^\ast$} 
{\begin {tabular} {ccccccccccc} \toprule
      && $T$   && $P$     && $V$        && $H$       && $S$\\
State && $[{\rm K}]$ && $[{\rm atm}]$ && $\rm [cm^3/g]$ && $[{\rm J}/{\rm mol}]$ && $[{\rm J}/ ({\rm mol}\, {\rm K})]$\\
\cline{1-1} \cline{3-3} \cline{5-5} \cline{7-7} \cline{9-9} \cline{11-11}
1     && 300       && 1           && 1.004        && 2040.8     &&  7.1\\
2     && 330       && 1           && 1.016        && 4300.2     && 14.3\\
3     && 330       && 100         && 1.011        && 4452.9     && 14.2\\
4     && 300       && 100         && 0.999        && 2206.5     &&  7.1\\ \bottomrule
\end{tabular}}
\tabnote{Refer to \cite{Chase1998}
}
\label{table.08}
\end{table}

\bibliographystyle{tfo}
\bibliography {references}

\end{document}